\documentstyle[12pt]{article}
\input{psfig}
\begin{document}
\baselineskip 23pt
\bibliographystyle{unsrt}
\pagestyle{plain}

\title{From Neurons to Brain: Adaptive Self-Wiring of Neurons}
\author{Ronen Segev and Eshel Ben-Jacob\\
School of physics and astronomy,\\
Raymond \& Beverly Sackler Faculty of exact Sciences,\\
Tel-Aviv University, Tel-Aviv 69978, Israel \\}
\maketitle








\begin{abstract}
During embryonic morpho-genesis, a collection of individual neurons
turns into a functioning network with unique capabilities. Only
recently has 
this most staggering example of emergent process in 
the natural world, began to be studied. Here we propose
a navigational strategy for neurites growth cones, based
on sophisticated chemical signaling. We further propose that the embryonic
environment (the neurons and the glia cells) acts as an excitable media
in which concentric and spiral chemical waves are formed. Together with
the
navigation strategy, the chemical waves provide a mechanism for 
communication, regulation, and control required for the adaptive self-wiring
of neurons.
\end{abstract}

\section{Introduction}

The brain is probably one of the most challenging and alluring complex system
scientists can study \cite{Abeles91}. And indeed, much effort has been
devoted to brain  
studies from the physiological level of ionic channels to the
philosophical level where 
questions about intelligence, self-awareness, and conciseness are
discussed. Yet, one of the 
fascinating aspects about the brain has almost been completely ignored until
recently. We refer to the process in which a collection
of individual neurons are transformed into a functioning network with unique 
capabilities-the brain. This emergence process can not be totally
determined by the 
stored genetic information. In a human brain, for instance, there are 
approximately $10^{11}$ neurons that form a network with more that $10^{15}$
synaptic connections. The precise structure of such a network can not be 
stored genetically. The human $DNA$ is composed of about $10^9$ bases, so it 
lacks sufficient memory for the detailed structure of a brain.
The alternative extreme explanation, of total randomness, could not
be correct as well. After all, we know that while on the micro level (up to
about $1mm$) the structure appears to be random, on the macro level
(above $1cm$) the brain's structure is quite deterministic.
In addition, the brain structure varies from species to 
species, and, within a given species, some brain skills are inheritable
(vs. learned). So clearly some elements of the brain structure must
be dictated by stored genetic information. 

At present we do not know exactly which 
information of the brain structure is stored, and the general question 
about the role of randomness vs. determinism during the emergent process is
still open. 

We describe here elements of a
novel strategy for the emergent process involving interplay between
randomness and determinism (via stored genetic information). Our
proposed model is  
for simplified $2D$ systems. As such, it is far from being a full description
of the brain adaptive self-wiring. Yet, if tested and shown to be
correct, it will  
provide an important step towards understanding the emergence process
in a real brain.

A major obstacle in unraveling the fundamental principles of adaptive 
self-wiring is the complexity of the real brain. Hence much effort is 
devoted to in-vitro experimental studies of much simpler $2D$ systems
\cite{Dwir96,TL96}. In 
these experiments, neurons (primary or cell line) are placed on a 
Poly-L-Lysine (PLL) surface forming a simple $2D$ system. The latter
facilitates microscopic monitoring of the 
self-wiring and allows performing of parallel measurements of the
electrical activity. The theoretical studies presented here are for
such simpler $2D$ systems. 

Our model is composed of two main elements: 1. A navigation strategy for
the micro level, i.e. length scale below $1mm$ or about $10^3-10^4$ cells 
area in $2D$ systems. 2. A mechanism for adaptive self-organization on the
meso (about $1mm-1cm$) and macro scale (above $1cm$). 

The contemporary view is that the brain structure is essentially
deterministic on a large scale but probabilistic on a small scale
\cite{Abeles91}. As 
a consequence the network has no optimal structure. But we believe that
neural networks can be constructed in an optimal way, and this optimal
way is derived from the biological mechanisms that construct the
network. In any approach to the construction of the network there
must be a precise strategy by which the neurons find each other
to establish synaptic connections. At the beginning of the      
growth process the neurite has to migrate from it's own cell soma. The    
neurite migrates to the area in which it is supposed to form a synaptic   
connection. In this area there are many neurons, each of which is a
possible target cell for the neurite.
When the neurite approaches one of the possible
target cells, with which it will finally form a synaptic connection, 
it has to be attracted to that cell soma. 

The navigation strategy we propose
is based on sophisticated means of chemical signaling for communication and
regulations, including repulsive and attractive chemotaxis (chemotaxis
is a movement based by a gradient of a chemical agent.) The strategy
is described in section 2 and 3.

For reasons presented in section 5 we propose that the
additional mechanism for self-wiring on the meso and macro scales is
based on chemical waves. This proposal is inspired by the $cAMP$ waves
employed during aggregation of $Dictyostelium$ $discoideum$ amoeba
\cite{Newell78,KL93} and
is also motivated by the fact that chemical waves (including of
$cAMP$) play role during the brain activity.

\section{Neurite as Amoeba with a tail}

Neurites are equipped for navigation with a unit called growth
cone \cite{TN91}. It is known that the growth cone is
capable of measuring concentration and concentration gradients
 of chemicals \cite{TL96,TL91}, including 
the gradients of repulsive and attractive chemotactic agents.

The movement of the growth cones appears to be a non-uniform random walk
with the highest probability to move forward (``inertia'') and a high
probability to move backward (``retraction'') \cite{BR97,BU94}. The
typical growth rate 
is of the order of micron/minute \cite{GB79}. In the presence of a
chemotactic agent, the 
movement is biased towards (attractive chemotactic) or away (repulsive
chemotactic) from the chemical gradient. The movement is reminiscent
of that of amoebae Dictyostelium discoideum \cite{Newell78} in the presence of
chemotactic materials. For that 
reason R. Lumins refers to the growth cone and its neurite as ``amoeba
with a tail'' \cite{Lumins97}.

In modeling the neurite navigation we were inspired by the bions
model used in the study of amoebae aggregation \cite{KL93} and by the
communicating 
walker model used in the study of bacterial colonies \cite{EB95,EB94,EBCP97}.
In the model the growth cones are represented
by walkers which perform off-lattice non-uniform random
(biased) walk \cite{SB98}. The chemical dynamics (e.g chemotactic agents,
triggering field) are described by continuous
reaction-diffusion equations solved on a tridiagonal lattice with a
lattice constant $a_0=10\mu m$.
Each of the soma is represented by a stationary (not moving) unit
occupying one lattice cell. The neurite are simply defined as the
trajectory performed by the growth cone.

\section{Chemotactic Navigation}

We assume that each of the soma cells continuously emits a repulsive
agent whose concentration is denoted by $R$. In the model, $R$ satisfies
the following reaction diffusion equation:
\begin{equation}
   \frac{ \partial R}{ \partial t}=D_R \nabla^2 R +
   \Gamma_{R} \sum_{{ }^{soma}} \delta ( \vec{r} - \vec{r_j} )
   -\lambda_R R
\end{equation}
Where $D_R$ is the diffusion coefficient, $\lambda_R$ is the spontaneous
decomposition rate and $\Gamma_R$ is the emission rate by the soma cells. 

When a neurite first sprouts it is mainly affected by the repulsive agent
and moves away from its ``mother'' soma cell. It then continues to
move on a trajectory which maximizes the distances from the surroundings
soma cells (Fig 1a).

When a neurite reaches a specific length determined by its soma cell
in a manners described below, it does two things: 1. It switches
its chemotactic sensitivity from sensitivity to the repulsive
agent to sensitivity to the attractive. 2. It emits a quantum
of a triggering material (which satisfies an equation similar to
eq. 1. Soma cells in the neighborhood respond by
emitting a quantum of
attractive agent if they sense an above threshold concentration of the
triggering material. As a result, the growth cone moves towards the
soma cell with the strongest attractive response (typically, the one
closest to the growth cone, see Fig 1b). The above features are included in the
model, as we show in detail in Ref \cite{SB98}.

\section{Simulation of the chemotactic navigation}

As was mentioned, the reaction-diffusion equations are solved on a
tridiagonal lattice with a  
lattice constant \( a_0=10\mu m \). Thus, the fact that
the soma occupies one lattice cell is in   
agreement with their typical size. The typical  
size of the simulated system is about \( 200-400 \) \( a_0 \) and the
distance between cells is about $25a_0$. A typical
diffusion coefficient \( D \) of the chemicals is of the 
order of \( 10^{-6}-10^{-7} cm^2/sec \) \cite{GH98}. Time is measured
in units of 
$10sec$, thus the dimensionless diffusion 
coefficients are of the order of 10. 
 
In the simulations, the walkers growth rate is about $1/64$ in
dimensionless units which is in agreement with the measured
growth rate. To demonstrate the
efficiency of our proposed chemotactic navigation strategy, we
consider two soma cells with a barrier in between (Fig 2). The cell on
the right is a ``normal'' soma cell, while the one on the left is a
``variance'' which is incapable of emitting neurites. The choice of only
one cell emitting neurites is in order to make the wiring pattern more
transparent. First, in the absence of chemotactic communication (Fig
2a) the barrier prevents the formation of synaptic connections between
the two cells. When included, the chemotactic communication enables to
over come the barrier effect and the two cells are wired as is shown
in Fig 2b.

We have mentioned that the growth cones` ``sensitivity switch'' (from
repulsive to attractive agent) is controlled by the soma. 
Actually, as is discussed in details in Ref \cite{SB98}, we propose
that the ``sensitivity switch'' is directly controlled by the
metabolic state of the growth cone which is indirectly controlled by
the soma. In the model, the metabolic state is represented by an
``Internal energy'', which is supplied at a given ``feeding'' rate by
the soma.
The rate of consumption of the internal energy is proportional to
the neurites' length. Thus, for a given rate of feeding the ``internal
energy'' will decrease for a given length. At
this length the growth cones switches its chemotactic sensitivity. It
means that the soma controls the ``sensitivity switch'' of the growth
cone via adjustment of its feeding rate. This way the soma can
control
the distance at which its various neurites form the synaptic
connections. An example is shown in fig 3. Here the central cell is
``normal'' and all the target cells are ``variance'' cells. The first two
neurites are fed at a low rate. Hence they are connected to the
nearest neighbors. The third neurite is fed at a higher rate. Thus it
forms a connection further away. 

\section{A need for additional mechanism}

So far we have described a navigation strategy for the microlevel.
This strategy on its own is limited in the variety and complexity of
the wiring  
patterns it can produce (see Fig 4). In these figures we see that the 
patterns have simple homogeneous structure of connectivity. 
Hence the genetic effect on the 
wiring patterns produced by this strategy has to be limited.
At the same time it is known that the brain has a specific organization on
the macro-scale (centers of activity over $1cm$) which is genetically
determined. In general it is not known what happens on the meso-scale 
between the apparent randomness of the 
micro-scale and the deterministic organization of the macro-scale 
\cite{Abeles91}. We believe (unlike Ref \cite{Abeles91}) that the
micro-structure is 
not entirely random nor the macro-structure is fully deterministic. 
But more important, we expect a gradual transition from the meso-structure 
to the macro-structure with partial deterministic organization on the 
meso-scale. If so, there should be an additional mechanism (to the navigation
strategy) to enable meso and macro scales self organization.

\section{The excitable media mechanism}

It is well established that chemical waves, which affect the electrical 
activity of neurons, can propagate through out the cortex
\cite{Mu89,Fernandes93,Ruppin97,Martins87,Gorelova83}.  
This brings to mind that such chemical waves can regulate the wiring 
process over the meso and macro scales. 

A classical example of self-organization mediated via chemical waves
is the aggregation of the slime mold Dictyostelium amoebae during
starvation. These microbes collectively form an excitable media which
produces spiral waves of $cAMP$. The movement of the individual
organism is biased by the local gradient of the $cAMP$ (chemotaxis
response). The result is regulated movement towards the center of the
colony along well defined stream lines \cite{Newell78,KL93}.

The embryonic media is composed of two main cell types: neurons and 
glia. The Glia cells are the majority of the cell population in the brain 
($~90\%$ of the brains cells are glia)\cite{TN91}. It is widely
accepted that the glia  
give the structural support to the brain and they play part the main in the 
neurons nutrition. Inspired by the aggregation of slime mold amoebae
we propose additional role to the glia: they acts as an excitable
media and form chemical waves of at least one additional agent refered
to as Glia Chemical Agent (GCA) needed for self-wiring on the meso and
macro scale.
The GCA (which can also be $cAMP$) plays a similar role to that of the
$cAMP$ during amoebae aggregation. It acts as an additional
chemotactic agent for the growth cone movement.

To model the glia role as excitable media we place randomly additional
stationary units 
(representing the glia) on the 2D hexagonal 
lattice with density $\rho_G$. Typically the $2D$ density   
$\rho_G$ should be $0.05-0.2$ in numbers. 
Each Glia unit can be in one of three possible states:
1. rest state, 2. excitable state, and 3. refractory state.  

When the unit is at rest state, at each 
time step, it measures the concentration of the chemotaxis at its 
location. Once the concentration is above a threshold level, the 
unit becomes excited (enter state 2) and starts to emit a quantum 
of GCA during $\tau_1$ time units. Then it enters the 
refractory state and it is immune to further excitation. After another
$\tau_2$ time units the state progresses to rest state.

Numerical experiment of these system reveal \cite{KL93} 
two basic patterns of waves that propagate in the medium. In  figure 5a we 
show concentric waves that propagate in response to a periodic pacemaker at
the center of the grid, and In figure 5b we show a spiral waves that
propagate in the media in response to appropriate initial conditions.

As we mentioned the length and time units of our simulation
corresponds to $10\mu m$ and $10s$ respectively. 
To test the consistency of the model we compare the velocity of waves
propagation in the simulation with the experimental one. 
The measured velocity of $cAMP$ waves is about $300\mu m/min$
\cite{Newell78} and the
measured velocity of the waves in the cortex is about $3000\mu m/min$
\cite{Gorelova83}.

The velocity of the waves in simulation is 
$10a_0$ per $1$ time step which corresponds to $10^{-2}cm/s$. This is
in agreement with the wave velocity measured in experiments
($\sim0.06 cm/s$).

We further assume that above a threshold level of the GCA concentration
it acts as an additional chemotactic agent (either repulsive or
attractive) on the movement of the walkers (growth cones). In Fig 6 we
demonstrate the GCA effect in a system of 100 cells. The soma are
restricted to emit only one neurite each in order to make the wiring
patter more transparent to the eye. In Fig 6a there are no chemical
waves and indeed the wiring pattern ``looks'' random . Next we placed a
pacemaker at the center to produce concentric waves. 
In Fig 6b we took the GCA to act as an attractive chemotaxis agent,
and in Fig 5c to act as a repulsive one. 

Finally in Fig 7 we demonstrate the effect of spiral wave. In this
case the system is composed of $1250$ cells, each soma emits two
neurites and repulsive response is taken.
In order to make the pattern of the network more transparent to the
eye we line up neurites, i.e we draw a line between the cell soma of
and the walkers position (fig 8). 
\section{Conclusions}

We have presented a navigation strategy for micro-level network
organization and an excitable media based wiring mechanism for the
meso and macro levels organization. These mechanisms leads to the
formation of neural networks with 
different structures, which can be genetically dependent. Our results
raise two fundamental issues: 1. One needs to develop characterization
methods (beyond number of connections per neurons) to distinguish the
various possible networks. 2. The relations between the network
organization and its computational properties
and efficiency. 

To clarify what we mean by characterization method, we consider the
following example in which the network is mapped into a directed graph. 
Each neuron is represented by a vertex and each
synaptic connection between neurons is represented by a directed edge
between the two 
vertices representing the neurons. We can map the graph structure into
an adjacency matrix where  
there is 1 in the i,j entry of the matrix if there is a directed edge between
the i and j vertices and 0 otherwise. Then using an algebraic
method, such as spectral theory of matrices,
we can analyze the adjacency matrix of the graph which characterize
the network. 
In other word, just as astronomers study spectra to determine the
make-up of distance stars, we try to deduce the principle properties
and structure of graph from its adjacency matrix spectrum \cite{Chung97}. 

The next step of the endeavor presented here is to include the effect
of the network 
electrical activity on its self-wiring. Once some initial synaptic
connection are formed, the network begins its electrical activity. It
is natural to expect that from this point on the wiring dynamics depend
on the electrical activity. An example to such a dependence is to
assume that the neurons emit chemoattractant while it bursts a train
of spikes. In addition we assume that a growth cone is more sensitive
to the chemoattractant when its soma bursts. Using such 
navigation strategy we expect that neurons with correlated electrical
activity will have higher probability to form a synaptic connection,
as is shown in fig 9.    

To conclude, we are still far from understanding the emergence of a
functioning brain from a collection of neurons. Yet we believe our
studies provide a useful first step towards this goal.

\section*{Acknowledgments}

The research was supported in part by grant No. BSF 92-00051 from the
Israel USA binational foundation, by grant No. 593-95 from the Israeli
Academy of Science, the Siegl Prize for Research, Sackler institute
and The Adams Super Center.

\section*{Figure1}

\section*{Figure2}
the effect of turning off the chemotactic communication. a. When we
turn off the chemotaxis communication the walker are unable to reach
their synaptic target. b. When we turn the communication on the walker
can migrate around the barrier.

\section*{Figure3}
Simulations of a system composed of 30 cells. Only the cell at the center
is "normal" and all other cells are "variance". The central cell has
four nearest neighbor (NN) cells and eight next nearest neighbor (NNN) cells.
At the beginning of the growth the feeding rate is low. 
Thus the central cell is wired
only to its NN cells. After The central cell forms two connections the
"feeding" rate doubles. The new neurites navigate to the NNN cells. It
demonstrates the manner in which the soma cell can regulates self-wiring.

\section*{Figure4}

\section*{Figure5}
Waves propagating in the media for $\rho_B=0.25$, $\tau_1=.8$,
$\tau_2=5$, $D=10$, and $C_T=.001$.
a. Spiral wave propagating in a 2D excitable media on a $200\times200$  
grid.
b. Target waves propagating in response to a pace make at the center
of the grid (again $200\times200$).

\section*{Figure6}
In each network there are $100$ cells which are restricted to emit
only one neurite. a. There is no wave source at the center. The
resulting network ``looks'' random. b. There is a pacemaker at the
center of the grid. The GCA is assumed to be a repulsive agent. As a result
the neurite migrate out from the center. c. Again a pacemaker at the
center but now the GCA is assumed to be an attractive agent.

\section*{Figure7}
The effect of spiral wave on the network structure. In the simulation
we assumed the GCA is an attractive agent. The spiral wave propagates
in response to appropriate initial conditions.

\section*{Figure8}
The same network as in figure 7 but now the final position of the
growth cones is linked by a straight line in order to make the
connections pattern more transparent.

\section*{Figure9} 
Example of the effect of the correlation in the electrical
activity. The left and right neurons fire simultaneously and the
first connection is formed between them. We remove the walkers which
failed to create a synaptic connection in order to make the
connections more transparent.

\end{document}